\documentclass[prb,twocolumn]{revtex4}

\usepackage{graphics}
\usepackage{epsfig}

\begin{document}

\title{Self-consistent local GW method: Application to $3d$ and $4d$ metals}
\author{K. D. Belashchenko}
\altaffiliation{Present address: Department of Physics and Astronomy, University of
Nebraska, Lincoln, Nebraska 68588.}
\author{V. P. Antropov}
\author{N. E. Zein}
\altaffiliation{Permanent address: Russian Research Center ``Kurchatov Institute'', Moscow
123182, Russia.}
\affiliation{Ames Laboratory, Ames, Iowa 50011}

\begin{abstract}
Using the recently developed version of the GW method employing
the one-site approximation and self-consistent quasiparticle basis
set we calculated the electronic structure of $3d$ and $4d$
transition metals at experimental atomic volumes. The results are
compared with traditional local density approach and with
experimental XPS and BIS spectra. Our results indicate that this
technique can be used as a practical starting point for more
sophisticated many-body studies of realistic electronic
structures.
\end{abstract}

\maketitle


\section{Introduction}

The electronic properties of many solids are described reasonably
well in the local density approximation (LDA) of the density
functional theory (DFT),\cite{DFT} which is essentially a
single-particle theory representing a cardinal simplification of
the original many-body problem. However, the LDA becomes
inadequate in materials exhibiting strong spatial correlations
between different electrons in atoms with localized orbitals,
temporal correlations (retardation) associated with Coulomb
screening, etc. The corresponding difficulties are extremely hard
to overcome within the DFT without sacrificing the
first-principles approach.

One of the methods used to correct the deficiencies of LDA in the
description of excited states in metals and semiconductors is the
GW approximation (GWA).\cite{Hedin,AG-rev} The GWA usually
improves the band gaps in semiconductors and
insulators;\cite{AG-rev} in metals it may provide information on
the quasiparticle lifetimes and renormalization which is absent in
the DFT.\cite{AG-rev,Echenique} However, in all treatments used
until recently, the GWA was employed in a non-self-consistent
fashion, by using unrenormalized Green's functions based on the
Kohn-Sham wave functions obtained in LDA. Such an approach is
easier to implement, but it is internally inconsistent, does not
correspond to any Luttinger-Ward functional\cite{LW} and hence
violates basic conservation laws,\cite{Baym} and produces
different results depending on the approximation used to solve the
Kohn-Sham equations. On the other hand, studies of the homogeneous
electron gas have shown that GW self-consistency worsens the
agreement with experiment at typical metallic
densities,\cite{Holm} highlighting the limitations of GWA which is
strictly correct only in the high-density limit.

Recently, two self-consistent realizations of the Green's function
method were demonstrated. One of them\cite{ZA} was tailored for
transition metals and employed the one-site approximation for the
self-energy which is based on the assumption that dynamical
screening in $d$-metals is well described at the intraatomic
level. Some spin-selective diagrams beyond the GW set were also
included. The results for Fe and Ni were quite reasonable,
although no improvement was obtained compared to LDA. While the
choice of GW diagrams is unjustified for the homogeneous electron
gas at typical metallic densities and likewise for semiconductors
(the diagrams that are left out do not contain any small
parameter), it was noted that the situation may be better in
transition metals, because high orbital degeneracy provides an
additional enhancement to the diagrams with the largest number of
closed electron loops favoring the GW set.\cite{ZA} In addition,
for transition metals the ``local'' (one-site) approximation for
the self-energy is quite reasonable due to the short screening
length and greatly facilitates self-consistent calculations.

Another realization\cite{KE} using GW set with the full
$\mathbf{k}$-dependence of the self-energy was applied to
elemental semiconductors. It was found that self-consistency and
accurate treatment of core electrons improve the agreement with
experiment for the band gaps in Si and Ge.

While the adequacy of GWA for the studies of metals and
semiconductors has not been firmly established from the point of
view of the many-body theory, this method may be regarded as one
of the possible steps toward a consistent Green's function-based
scheme. Therefore, it is important to establish the accuracy of
this approximation for different materials, and in particular for
transition metals where, as noted above, there are reasons for it
to work better than in the homogeneous electron gas. To this end,
in this paper we calculate the conduction-band spectra for all
elemental $3d$ and $4d$ metals using the self-consistent GW
approach of Ref.~\onlinecite{ZA} and compare them with the
observed ones and with those given by the LDA.

\section{Self-consistent GW method\label{method}}

The technique used in this paper was introduced in Ref.~\onlinecite{ZA}.
Here we describe some points in more detail.

The self-energy $\Sigma =\delta \Phi /\delta G$ is obtained from the
Luttinger-Ward generating functional $\Phi $ defined by the set of skeleton
graphs.~\cite{LW} The Hartree diagram gives the local contribution $V_{%
\mathrm{H}}(\mathbf{r})$, and the exchange diagram contributes $\Sigma_x=V(%
\mathbf{r}-\mathbf{r}^\prime)\int\mathrm{Im}G(\mathbf{r},\mathbf{r}%
^\prime,\epsilon)d\epsilon/\pi$ where $V$ is the Coulomb
potential. The total contribution of the remaining GW sequence
(the ``correlation term'') is
\begin{eqnarray}
\Sigma_{c}(\mathbf{r},\mathbf{r}^{\prime },\epsilon)&=&-\int G(\mathbf{r},%
\mathbf{r}^{\prime},\epsilon-\omega)V(\mathbf{r}-\mathbf{r}_{1})  \nonumber
\\
&\times &\Pi (\mathbf{r}_{1},\mathbf{r}_{2},\omega )W(\mathbf{r}_{2},\mathbf{%
r}^{\prime },\omega )\,d\mathbf{r}_{1}\,d\mathbf{r}_{2}\,\frac{d\omega }{%
2\pi i}  \label{sigmac}
\end{eqnarray}
Here $W$ is the effective (screened Coulomb) interaction defined by the
integral Dyson equation
\begin{equation}
W=V-\int V(\mathbf{r}-\mathbf{r}_{1})\Pi (\mathbf{r}_{1},\mathbf{r}%
_{2},\epsilon )W(\mathbf{r}_{2},\mathbf{r}^{\prime },\epsilon )d\mathbf{r}%
_{1}d\mathbf{r}_{2}  \label{W}
\end{equation}
and the polarization operator
\begin{equation}
\Pi (\mathbf{r},\mathbf{r}^{\prime },\omega )=\int G(\mathbf{r},\mathbf{r}%
^{\prime },\epsilon )G(\mathbf{r}^{\prime },\mathbf{r},\epsilon +\omega
)d\epsilon .  \label{Pi}
\end{equation}%
Contour of integration in Eqs.~(\ref{sigmac}),(\ref{Pi}) is directed along
the imaginary axis and embraces the cut on the real axis from the Fermi
energy $E_F$ to the external energy. The GW approximation may be modified by
the inclusion of vertex corrections in the polarization operator (\ref{Pi})
which may improve the results in some cases.

The calculations are drastically simplified by the use of the
one-site approximation~\cite{ZA} (OSA), in which the self-energy
is calculated only on one lattice site neglecting all non-diagonal
maxrix elements connecting different lattice sites. Thus, in OSA
the self-energy depends on energy and on the coordinates
$\mathbf{r}$, $\mathbf{r}^\prime$ belonging to the same unit cell.
In order to implement OSA we first need to choose an appropriate
on-site basis. Inspired by the good description of the band
structure of densely packed solids achieved in the atomic sphere
approximation within the linear muffin-tin orbital method
(LMTO-ASA), we used a very similar, minimalistic approach. For the
basis set we also use just one function per each angular momentum
$l$ and its projection $m$ (below we denote $L\equiv lm$), along
with its energy derivative. The radial basis functions $\phi_l(r)$
satisfy the ``Schr\"odinger-type'' equation
\begin{equation}
\left[\epsilon_l+\frac{\Delta_l}{2}-V_{\mathrm{H}}-\widehat\Sigma_l(%
\epsilon_l)\right] \phi_l(r)=0  \label{Schroed}
\end{equation}
where $\Delta_l$ is the radial part of the Laplasian, and
$\widehat\Sigma_l$ is an integral radial operator whose kernel is
obtained from $\mathop{\mathrm{Re}}\widehat\Sigma$ by projecting
onto the $l$ subspace:
\begin{equation}
\Sigma_l(r,r^\prime,\epsilon) =\frac{1}{2l+1}\sum\limits_m\int Y_L(\hat r)%
\mathop{\rm Re}\Sigma_{xc}(\mathbf{r},\mathbf{r}^\prime,\epsilon)Y_L(\hat
r^\prime)dodo^\prime
\end{equation}
where we denoted $\Sigma_{xc}\equiv\Sigma_{x}+\Sigma_{c}$, $Y_L$
are the spherical harmonics, and integration
is over the directions of $\mathbf{r}$,$%
\mathbf{r}^\prime$. The operator $\widehat\Sigma_l$ may be
represented as the sum of a local part $\Sigma_l^{(l)}(r)$ similar
to an external potential and a nonlocal part
$\widehat\Sigma_l^{(n)}$ which (after acting on $\phi_l$) gives a
linear combination of $\phi_{l^\prime}$ with $l^\prime\ne l$.

Similarly to the LMTO-ASA method, the solutions of
Eq.~(\ref{Schroed}) are only found at one energy $\epsilon_l$ for
each $l$ which is fixed at the center of gravity of the
corresponding band. We may safely discard the imaginary part of
the self-energy operator because it is small in the vicinity of
$E_F$ where $\epsilon_l$ is usually chosen. Thus, our radial basis
functions are real. The non-local equation (\ref{Schroed}) is
solved by iterations. If $\phi_l^n$ is the solution after $n$
iterations, then the solution on iteration $n+1$ is obtained using auxiliary functions $%
f_l$ and $g_l$ defined as
\begin{eqnarray}
\left[\epsilon_l+\frac{\Delta_l}{2}-V_{\mathrm{H}}-\Sigma_l^{(l)}(r)\right]
f_l(r)&=&
\widehat\Sigma_l^{(n)}(\epsilon_l)\phi_l^n, \\
\left[\epsilon_l+\frac{\Delta_l}{2}-V_{\mathrm{H}}\right] g_l(r)&=&0
\end{eqnarray}
according to $\phi_l^{n+1}(r)=f_l(r)+A\,g_l(r)$, where the
coefficient $A$ is found by normalizing $\phi_l^{n+1}$. Just as in
the LMTO method, in addition to $\phi_l$ we compute its energy
derivative $\dot\phi_l\equiv\partial\phi_l/\partial\epsilon$:
\begin{equation}
\left[\epsilon_l+\frac{\Delta_l}{2}-V_{\mathrm{H}}-\widehat\Sigma_l(%
\epsilon_l)\right] \dot\phi_l=-\phi_l+\frac{\partial\widehat\Sigma_l(%
\epsilon_l)}{\partial\epsilon_l}\phi_l  \label{Schroed1}
\end{equation}

To stabilize the solution of the Schr\"odinger equation (\ref{Schroed}), we
added the exchange with the nearest-neighbor cells to $\Sigma_{xc}$. The
corresponding matrix element was subtracted from Green's function (\ref%
{GF}) below.

With the on-site self-energy
$\Sigma_{xc}(\mathbf{r},\mathbf{r}^\prime,\epsilon)$ defined for
$r,r^{\prime}$ within the single unit cell we can calculate its
matrix elements between the muffin-tin ``eigenfunctions''
$\chi_{\mathbf{k}\nu}$, which in turn are the linear combinations
of $\phi_l$ and $\dot\phi_l$. Due to the $\mathbf{k}$-dependence
of $\chi_{\mathbf{k}\nu}$ the self-energy also acquires
$\mathbf{k}$-dependence, just as the local potential in LDA.
Finally, on-site Green's function is found by integration using
the tetrahedron method:
\begin{equation}
G({\bf r,r'},\varepsilon )= \sum_{{\bf k}\nu }
\frac{\chi_{\mathbf{k}\nu}^{R}({\bf
r})\chi_{\mathbf{k}\nu}^{L*}({\bf r'}) } {\varepsilon -
\lambda_{\mathbf{k}\nu}} \label{GF}
\end{equation}
where $\chi_{\mathbf{k}\nu}^{R}$ and $\chi_{\mathbf{k}\nu}^{L}$
are right and left eigenvectors, and $\lambda_{\mathbf{k}\nu}$ the
eigenvalues of the non-Hermitian operator $H_0+\Sigma_{xc}$
where $H_0$ is the Hartree Hamiltonian. 
Eq. (\ref{GF}) imposes the locality condition and is equivalent to
the self-consistency condition of the dynamical mean-field
theory.\cite{DMFT}

\section{Conduction band spectra of transition metals\label{spectra}}

Using the technique described above (below referred to as ``GW'') we
calculated the spectral densities $N(\epsilon)=\mathop{\mathrm{Tr}}%
\mathop{\mathrm{Im}}G(\epsilon)$ for elemental $3d$ and $4d$
metals, where LDA has long been the only appropriate
approximation. We used the fcc structure for hcp elements (Sc, Ti,
Y, Zr, Tc, Ru, Co) and the bcc structure for Mn. All atomic
volumes were taken from experiment. Cr was treated within the
single bcc cell, and hence was non-magnetic.

The iterations were started from the LDA potential, and at each
iteration the non-local self-energy was mixed to this potential
with a gradually increasing weight, so that in the end only
$\Sigma_{xc}$ was left. In the final state the magnitude of
$\Sigma_{xc}$ differs by about 40\% from its initial value based
on the LDA wave functions. The iterational procedure is rather
stable in all metals except Ni where the magnetic moment is very
sensitive to the details of the calculation.

The results are shown in Figs.~1,2 along with the LDA densities of states
(DOS) and the experimental XPS and BIS spectra taken from Refs.~%
\onlinecite{XPS,BIS}. Strictly speaking, comparison with experiment requires
the calculation of the corresponding matrix elements, but we believe that in
the present context some qualitative conclusions can be drawn based on the
spectral densities.

The main differences between the LDA DOS and GW spectral density
may be summarized as follows. The conduction band widens as all
DOS features are pushed away from $E_F$; this outward shift is
roughly proportional to the distance from $E_F$. Moreover, all DOS
features are increasingly smeared out due to the decreasing
quasiparticle lifetime as the distance from $E_F$ increases.
Substantial spectral weight is transferred from the
quasiparticle states to the incoherent ``tail'' extending far below $%
E_F $.

The LDA DOS for $d$-metals is typically too small to account for
experimentally measured electronic specific heat $C=\gamma T$. As seen from
Figs.~1,2, the GW spectral density at $E_F$ is generally smaller compared to
LDA. Although to obtain $\gamma$ we have to remove the renormalization
factor $Z$ from $\mathop\mathrm{Im}G$, the GW method does not improve the
overall agreement with $\gamma$ measurements.

\begin{figure*}[pbt]
\epsfig{file=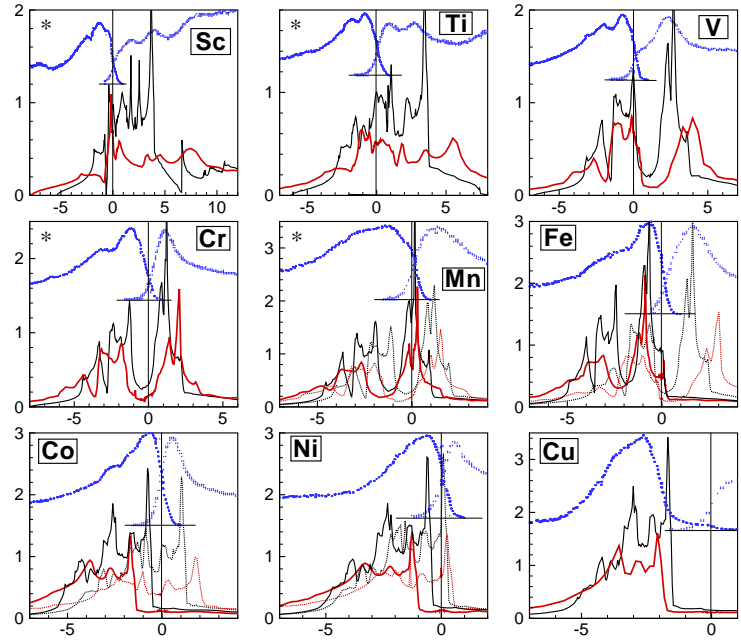,width=0.54\textwidth,clip} \caption{Spectral
densities obtained in GW-OSA (red lines), LDA densities of states
(black lines), and experimental XPS and BIS spectra (blue dotted
lines shifted up) for $3d$ metals. For magnetic elements the
calculated curves for majority-spin and minority-spin electrons
are plotted in solid and dotted lines, respectively. Elements for
which the crystal structure was not experimental (including Cr)
are marked by a star in the upper left corner of the graph. In all
graphs the $x$ axis shows energy referenced from $E_F$ in eV, and
the $y$ axis denotes the spectral density in eV$^{-1}$ for the
calculated curves.}
\end{figure*}

\begin{figure*}[pbt]
\epsfig{file=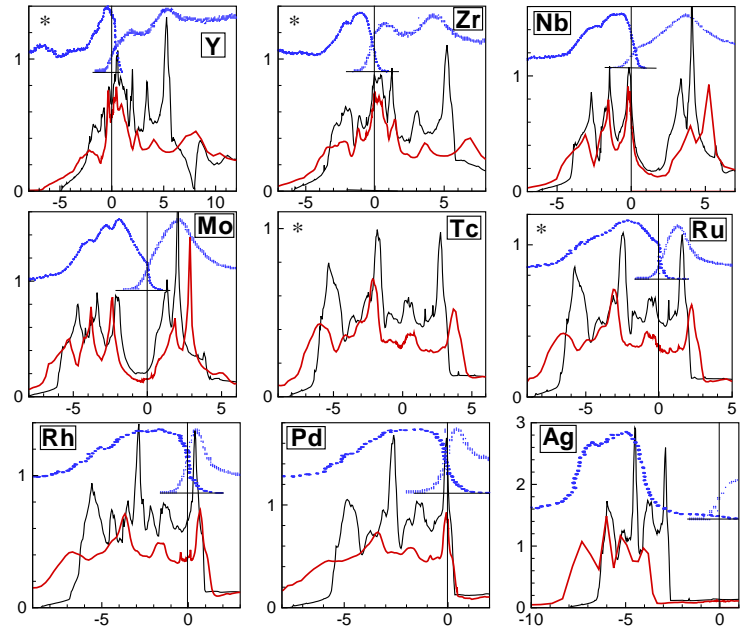,width=0.54\textwidth,clip} \caption{Same as
Fig.~1 but for $4d$ metals.}
\end{figure*}

The differences between the GW and LDA spectra are most notable
for early transition metals (treated here in the fcc structure)
where the strongest unoccupied peaks given by LDA are almost
completely smeared out in GW. Notably, the worst agreement with
experiment for the LDA spectra is observed for the same elements.
In particular, the position of the Fermi
level obtained in LDA for hcp Sc and Y appears to be off by nearly 1~eV.\cite%
{XPS}

The agreement between LDA and GW improves as we progress to later
transition metals and as the fcc structure is replaced by the bcc
one. Already in V and Nb the GW and LDA curves are quite similar
except for the shift of the unoccupied $d$-peak by 1.5--2~eV to
higher energies.

As in LDA, the $3d$ metals from Mn to Ni were magnetic in our
calculation. For simplicity, we used the bcc structure for Mn, and
also the fcc structure for Co (this structure is often realized in
thin films). As it is seen from Fig.~1, the general features of
the GW spectral density described above are observed in these
metals as well. In general, the GW description also gives a
reasonable exchange splitting and magnetic moment. We obtained the
moments of 0.9~$\mu _{B}$ for Mn compared to 1.03~$\mu _{B}$ in
LDA, 2.3~$\mu _{B}$ for Fe compared to 2.25~$\mu _{B}$ in LDA, and
1.85~$\mu _{B}$ for fcc Co compared to 1.62~$\mu _{B}$ in LDA.
From these results it is clear that there is no systematic trend
for GW to give larger or smaller magnetic moments compared to LDA.
The magnetic moment in Ni is rather sensitive to various details
of the calculation, and proper convergence turned out to be
problematic. We believe that the approach based on Matsubara
Green's functions is necessary to stabilize this problem. Apart
from the exchange splitting, the shape of the spectrum is quite
stable. We also note that the magnetic moment is expected to be
sensitive to the choice of the skeleton graph set.

As an example of the general trend of band dilatation off the
Fermi level, the distance from $E_F$ to the upper edge of the
fully occupied $d$-band in Cu and Ag is notably larger in GW
compared to LDA, which results in a better agreement with
experimental XPS spectra. On the other hand, we also observe a
rather strong upward shift of the unoccupied peak in V, Nb, Fe and
Mo in obvious disagreement with the BIS spectra. This shift is
more pronounced compared to the downward shift of the occupied
states at a similar distance from $E_F$.

The results presented above demonstrate that self-consistent GF
approach with one-site approximation can be used for ground state
studies. For $3d$ and $4d$ systems we have shown that the GW
spectral density is generally similar to the LDA density of
states, while the GF approach includes typical Fermi-liquid
effects such as finite quasiparticle lifetime and self-consistent
renormalization. The preliminary comparison with experiment is
rather satisfactory and clearly indicates the problems that need
improvement beyond our simple GW-OSA technique are similar to
those in LDA --- the exchange splitting in ferromagnets, value of
$N(E_{F})$, and the unoccupied peak position which is too high for
some metals. In general, the presented technique based on the
Luttinger-Ward functional is a practical alternative to other
methods based on the density functional theory and can serve as a
reliable starting point for more sophisticated methods..

This work was carried out at the Ames Laboratory, which is
operated for the U.S. Department of Energy by Iowa State
University under Contract No. W-7405-82. This work was supported
by the Director for Energy Research, Office of Basic Energy
Sciences of the U.S. Department of Energy. NEZ acknowledges
support from the Council for the Support of Leading Scientific
Schools of Russia under grant NS-1572.2003.2.

\end{document}